\documentclass[aps,prd,twocolumn,superscriptaddress,floatfix,nofootinbib]{revtex4-2}

\usepackage{graphicx}
\graphicspath{{figures/}}
\usepackage{amsmath}
\usepackage{amssymb}
\usepackage{hyperref}
\usepackage{xcolor}
\usepackage{multirow}

\newcommand{\sqrts}{\sqrt{s_{\mathrm{NN}}}}

\newcommand{\muB}{\mu_B}
\newcommand{\avg}[1]{\langle #1 \rangle}
\newcommand{\gs}{\gamma_s}
\newcommand{\gq}{\gamma_q}

\newcommand{\fcore}{f_{\mathrm{core}}}

\newcommand{\chindf}{\chi^2/\mathrm{ndf}}

\begin{document}

\title{Strangeness, entropy, and chemical non-equilibrium from p+p to Pb+Pb: the $K^+/\pi^+$ horn on the system-size axis}

\author{Neeraj}
\email{neeraj.neeraj@cern.ch}
\author{Amal Sarkar}
\email{amal.sarkar@cern.ch}
\affiliation{Indian Institute of Technology Mandi, Kamand,
Himachal Pradesh, India}

\date{\today}

\begin{abstract}
The maximum (``horn'') of the $\avg{K^+}/\avg{\pi^+}$ excitation function in central Pb+Pb collisions near $\sqrts = 7.6$~GeV is interpreted as the competition between strangeness production and the faster rise of entropy production at the onset of deconfinement. Statistical hadronization analyses by Letessier, Rafelski, and collaborators have shown that the horn is described quantitatively only if the light-quark phase-space occupancy $\gq$ is free, and that $\gq$ then lies near the pion condensation limit from 30$A$~GeV upward, as expected when an entropy-rich deconfined source hadronizes suddenly. We carry this analysis along the system-size axis provided by NA61/SHINE (p+p, Be+Be, Ar+Sc, preliminary Xe+La) and NA49 (Pb+Pb). Our main new result is the first non-equilibrium analysis, to our knowledge, of the Be+Be and Ar+Sc $4\pi$ multiplicities, published in 2021 and 2024: in Ar+Sc, chemical equilibrium of light quarks ($\gq = 1$) is disfavored, but through the $K^+/K^-$ asymmetry rather than the pion yields, and no single thermal source describes the data, while the Be+Be data do not discriminate. For central Pb+Pb, our fits to the final NA49 data with the full hadron spectrum confirm the earlier result: $\gq = 1$ is disfavored from 30$A$~GeV upward, most robustly at 30$A$--80$A$~GeV, with $\gq$ at the pion condensation limit, while at 20$A$~GeV $\gq$ is not constrained. Varying one element of the analysis at a time, we show that this preference rests on the excess of produced hadrons per net baryon and, for the pions, on their Bose--Einstein enhancement near the condensation limit, the two ingredients of the entropy-excess picture developed by the same authors. Decomposing $\avg{K^+}/\avg{\pi^+}$ into strangeness and pion yields per wounded nucleon, we find that the step between Be+Be and Ar+Sc is carried by a doubling of strangeness per participant at nearly unchanged pion yield, whereas the horn appears only in Pb+Pb, where the pion yield per participant is suppressed at 20$A$--40$A$~GeV and then rises faster than in Ar+Sc. Parton--hadron transport with chiral symmetry restoration (PHSD) reproduces the Pb+Pb horn but predicts one also in Ar+Sc and Xe+La, where none is observed.
\end{abstract}


\maketitle

\section{Introduction}
\label{sec:intro}

Enhanced production of strange hadrons in nucleus--nucleus collisions, relative to
elementary interactions, was proposed as a signature of quark--gluon plasma (QGP)
formation because gluon fusion in a deconfined medium populates strange-quark phase
space far faster than hadronic reactions~\cite{Rafelski:1982pu,Koch:1986ud,Rafelski:2019twp}.
At the CERN SPS the NA49 collaboration found that the ratio of total yields
$\avg{K^+}/\avg{\pi^+}$ in central Pb+Pb collisions rises steeply with collision
energy, peaks near $\sqrts \approx 7.6$~GeV (30$A$~GeV), and then drops to a plateau:
the ``horn''~\cite{Afanasiev:2002mx,Alt:2007aa}.

Two pictures of the horn have been proposed. The Statistical Model of the Early Stage
(SMES)~\cite{Gazdzicki:1998vd}, formulated before the discovery of the horn and
reviewed in Ref.~\cite{Gazdzicki:2010iv}, assumes that the early stage of the
collision is in equilibrium and is described by a two-phase equation of state with a
first-order transition. The horn then marks the passage of the strangeness-to-entropy
ratio, measured by $\avg{K^+}/\avg{\pi^+}$, from its value in the confined phase to the
lower value of the deconfined phase, which carries many more thermal degrees of
freedom, through a mixed phase that extends over the onset energy range. The same onset
produces a change of slope (``kink'') in the pion yield per wounded nucleon and a
plateau (``step'') in the kaon inverse
slope~\cite{Alt:2007aa,Gazdzicki:2010iv,Grebieszkow:2012}. A mixed phase of this kind
implies a time-delayed expansion, which would show up as a long emission duration and a
large ratio of the outward to sideward pion interferometry radii~\cite{Rischke:1996em}.
This is not observed at the SPS or elsewhere~\cite{Lisa:2005dd}; in particular, the
NA49 radii in central Pb+Pb show no striking dependence on beam energy across the onset
region, 20$A$--158$A$~GeV~\cite{NA49:2007fqa}.

The kinetic picture explains the horn differently~\cite{Letessier:2002book}.
Strangeness is produced by gluon fusion and amplified by multiple collisions per
participant~\cite{Rafelski:1982pu,Koch:1986ud}, so that its yield per participant rises
rapidly with energy, and faster in large systems. The entropy (pion) yield of the
deconfined phase then rises faster still and overtakes it. An entropy per baryon well
above what an equilibrium hadron gas can hold was inferred from the particle
multiplicities of S-induced reactions at 200$A$~GeV already in Ref.~\cite{Letessier:1992xd}.
In sudden hadronization, without a long-lived mixed phase, this entropy excess is
carried into the hadrons out of chemical equilibrium: it appears as over-saturated
light-quark phase space, $\gq > 1$, and near the condensation limit as a Bose--Einstein
enhanced pion yield~\cite{Letessier:1999rm,Letessier:1998sz,Rafelski:2000by,Letessier:2000ay}. In statistical
hadronization fits of the full hadron yields, Letessier, Rafelski, and collaborators
found that $\gq$ jumps between 20$A$ and 30$A$~GeV from under-saturated values to values
near the pion Bose--Einstein condensation limit, $\gq \to e^{m_\pi/2T}$, together with
$\gs > 1$, i.e., that the hadronization dynamics changes between these two
energies~\cite{Letessier:2008,Rafelski:2008av}; the horn is described quantitatively
only with $\gq$ free~\cite{Letessier:2008}. In this picture the horn is a signature of
the dynamics of QGP hadronization, not only of its onset. The same non-equilibrium
hadronization, with $\gq$ near the condensation limit, describes the hadron yields in
Pb+Pb collisions at the LHC~\cite{Petran:2013lja}.

Fits that impose chemical equilibrium ($\gq = \gs = 1$, or $\gq = 1$ with a free
strangeness suppression factor $\gs$) have been used to describe the total yields in
central nucleus--nucleus collisions and to attribute the maximum of
$\avg{K^+}/\avg{\pi^+}$ to the energy dependence of the freeze-out temperature and
baryon chemical
potential~\cite{BraunMunzinger:2001mh,Andronic:2008gu,Andronic:2017pug,Becattini:2005xt}.
The quality of such fits is poor: for central Pb+Pb from 30$A$~GeV upward, our
equilibrium fits to the final NA49 data give $\chi^2 = 14$--25 for 6--8 degrees of
freedom (Sec.~\ref{sec:fits}).
The $\avg{K^+}/\avg{\pi^+}$ excitation function of a single heavy system does not by
itself separate these readings: each has enough freedom to follow it.

The system-size axis provides the missing lever arm. The NA61/SHINE experiment has
measured identified-hadron spectra and $4\pi$ multiplicities over the same energy range
in inelastic p+p~\cite{NA61SHINE:2017fne,Aduszkiewicz:2019vkj}, central
$^7$Be+$^9$Be~\cite{Acharya:2021ljw} and $^{40}$Ar+$^{45}$Sc~\cite{Adhikary:2024epjc}
collisions, with preliminary results for $^{129}$Xe+$^{139}$La~\cite{Panova:2026}.
Two features of these data are not yet explained: $\avg{K^+}/\avg{\pi^+}$ jumps
between Be+Be, which follows p+p, and Ar+Sc (the ``onset of
fireball''~\cite{Gazdzicki:2020,Larsen:2018}), yet Ar+Sc shows no horn. Statistical
hadronization analyses of the system-size dependence at the SPS covered C+C, Si+Si and
Pb+Pb at 158$A$~GeV~\cite{Becattini:2005xt} or assumed
$\gq = 1$~\cite{Vovchenko:2015idt,Motornenko:2018}. To our knowledge, the published
$4\pi$ multiplicities of Be+Be (2021) and Ar+Sc (2024) have not been analyzed with $\gq$
free, and model comparisons across the NA61/SHINE ladder have so far been
qualitative~\cite{Mackowiak-Pawlowska:2022,Panova:2026,NA61:2025strategy}.

In this work we use the system-size axis to test the chemical-equilibrium assumption
directly, rather than to assume it. We state at the outset what is new. For central
Pb+Pb the non-equilibrium analysis has been done
before~\cite{Letessier:2008,Rafelski:2008av,Rafelski:2003wh}; we repeat it with the final
NA49 data and the full PDG hadron spectrum, and it confirms the earlier result from
30$A$~GeV upward. The ingredients that, as we show, carry the preference for $\gq > 1$,
an entropy excess per baryon and Bose--Einstein enhanced pions near the condensation
limit, were identified in Refs.~\cite{Letessier:1992xd,Letessier:1999rm}. Our main new
result is the non-equilibrium analysis of the Be+Be and Ar+Sc multiplicities; new as well
are the decomposition of the horn into strangeness and entropy per participant along the
system-size ladder and a common $\chi^2$ comparison of transport and core--corona models
with the full ladder. Our analysis has three parts.
(i)~We decompose $\avg{K^+}/\avg{\pi^+}$ into strangeness and entropy (pion) yields
per wounded nucleon for p+p, Be+Be, Ar+Sc and Pb+Pb at matched energies. The
Be+Be$\to$Ar+Sc step is carried by a doubling of strangeness per participant at
nearly unchanged pion yield per participant, whereas the horn appears only in Pb+Pb,
where the pion yield per participant is suppressed at 20$A$--40$A$~GeV and catches up
at higher energies.
(ii)~We fit the $4\pi$ hadron multiplicities with the statistical hadronization model
using the full PDG hadron spectrum, with $\gq$ either fixed to 1 or free. In Ar+Sc,
$\gq = 1$ is disfavored, but the preference comes from the $K^+/K^-$ asymmetry rather
than from the pion yields, and no single thermal source describes the data; the Be+Be
data do not discriminate. For central Pb+Pb from 30$A$~GeV upward, the equilibrium
fits are disfavored relative to fits with $\gq$ at the pion condensation limit
($\Delta\chi^2 = 7$--15), as found by Letessier, Rafelski, and collaborators, most
robustly at 30$A$--80$A$~GeV; at 20$A$~GeV the data do not constrain $\gq$, so the
present data and analysis do not inform on the change of hadronization dynamics
between 20$A$ and 30$A$~GeV reported in Refs.~\cite{Letessier:2008,Rafelski:2008av}.
Varying one element of the analysis at a time, we find that the Pb+Pb preference does
not depend on the resonance spectrum or the data compilation; it requires the
net-baryon constraint and, for the pions, their Bose--Einstein enhancement near the
condensation limit.
(iii)~We confront hadronic (SMASH) and parton--hadron (PHSD) transport, a Glauber
core--corona model, and SMES with the full system-size ladder under a common $\chi^2$
definition. PHSD with chiral symmetry restoration reproduces the horn in
Pb+Pb but also predicts one in Ar+Sc and Xe+La, where the data show none.
Finally, we make explicit that $\avg{K^+}/\avg{\pi^+}$ determines only $\gs/\gq$, not
$\gs$, the combination quoted in the non-equilibrium
analyses~\cite{Letessier:1999rm,Rafelski:2003wh}, and that the double ratio
$(K^+/\pi^+)/(K^-/\pi^-)$ is insensitive to $\gs$ but sensitive to $\muB/T$ and $\gq$,
which fixes how these observables can be used to argue for or against equilibrium.

Throughout, results based on the preliminary Xe+La data~\cite{Panova:2026} are
marked as such and are not used in any calibration.
The paper is organized as follows. Section~\ref{sec:data} describes the data.
Section~\ref{sec:shm} defines the statistical hadronization model with $\gq$ and $\gs$.
Section~\ref{sec:perpart} presents the per-participant decomposition,
Sec.~\ref{sec:fits} the equilibrium and non-equilibrium fits, and
Sec.~\ref{sec:models} the model comparison along the system-size axis.
Section~\ref{sec:kminus} discusses $K^-/\pi^-$ and the double ratio,
Sec.~\ref{sec:discussion} what the presence or absence of the horn implies
for hadronization dynamics, and Sec.~\ref{sec:conclusions} summarizes.

\section{Experimental Data}
\label{sec:data}

All data used in this analysis are final, published results unless marked
preliminary, verified against the original publications and HEPData records where
available. Feed-down from weak decays is subtracted in the published NA61/SHINE, NA49
and STAR data. The AGS pion yields~\cite{Ahle:2000wq,Klay:2003zf} are not corrected for
it; they enter only the model comparisons of Sec.~\ref{sec:models}.

\paragraph*{Kaon-to-pion ratios.}
NA61/SHINE provides mid-rapidity and $4\pi$ $K^\pm/\pi^\pm$ ratios for inelastic p+p at
five beam momenta~\cite{Aduszkiewicz:2019vkj,NA61SHINE:2017fne}, central (0--20\%)
$^7$Be+$^9$Be at four to five~\cite{Acharya:2021ljw}, and central (0--10\%)
$^{40}$Ar+$^{45}$Sc at six~\cite{Adhikary:2024epjc}. Preliminary $4\pi$
$\avg{K^+}/\avg{\pi^+}$ for central $^{129}$Xe+$^{139}$La at four energies are digitized
from Fig.~5 of Ref.~\cite{Panova:2026} (Table~\ref{tab:xela_digitized}), with
statistical and systematic uncertainties taken from the inner and outer error bars.
Central (0--7.2\%) Pb+Pb data at five beam energies are taken from
NA49~\cite{Afanasiev:2002mx,Alt:2007aa}. Au+Au mid-rapidity ratios come from the STAR
beam energy scan (0--5\%, $\sqrts = 7.7$--39~GeV)~\cite{Adamczyk:2017iwn} and, below
$\sqrts = 5$~GeV, from the AGS compilation of Ref.~\cite{Alt:2007aa}
($4\pi$ and mid-rapidity)~\cite{Ahle:2000wq,Klay:2003zf}. The rapidity windows are
summarized in Table~\ref{tab:rapidity_windows}; mid-rapidity and $4\pi$ results are
never compared with each other.

\paragraph*{Hadron multiplicities and wounded nucleons.}
The statistical hadronization fits (Secs.~\ref{sec:shm}--\ref{sec:fits}) and the
per-participant analysis (Sec.~\ref{sec:perpart}) use the $4\pi$ multiplicities listed
in Table~\ref{tab:multiplicities}. For Pb+Pb these are the final NA49 results for
$\pi^\pm$ and $K^\pm$~\cite{Afanasiev:2002mx,Alt:2007aa}, $\phi$~\cite{NA49:2008goy},
$\Lambda$, $\bar\Lambda$, $\Xi^-$, $\bar\Xi^+$~\cite{NA49:2008ysv}, and
$\Omega$~\cite{NA49:2004irs}. The $\avg{\pi^+}$ at 40$A$--158$A$~GeV are those
recalculated in Ref.~\cite{Alt:2007aa}. At 158$A$~GeV the hyperon (10\% most central,
$\avg{W} = 335$) and $\Omega$ (23.5\%, $\avg{W} = 262$) yields are scaled to the 5\%
class of the pion and kaon data ($\avg{W} = 362$) in proportion to $\avg{W}$, with an
additional 5\% (hyperons) and 15\% ($\Omega$) systematic uncertainty for this scaling.
No final $4\pi$ antiproton yield is available for Pb+Pb. For p+p, Be+Be and Ar+Sc we use
the NA61/SHINE mean multiplicities of $\pi^\pm$, $K^\pm$ and $\bar p$ (and $p$ for
p+p)~\cite{NA61SHINE:2017fne,Acharya:2021ljw,Adhikary:2024epjc}. The mean number of
wounded nucleons $\avg{W}$ is taken from the same publications: for Be+Be and Ar+Sc
from the EPOS simulation with the experimental forward-energy selection, with
uncertainties of 0.8 and 1.2 that cover the spread of the alternative Glauber
estimates quoted there.

\begin{table}
\caption{\label{tab:rapidity_windows}
Rapidity windows of the kaon-to-pion ratios used in this work.}
\begin{ruledtabular}
\begin{tabular}{lccl}
System & Mid-rapidity & $4\pi$ & Source \\
\hline
p+p    & $|y| < 0.2$ & \checkmark & NA61~\cite{Aduszkiewicz:2019vkj} \\
Be+Be  & $y \approx 0$ (fitted) & \checkmark & NA61~\cite{Acharya:2021ljw} \\
Ar+Sc  & $|y| < 0.2$ & \checkmark & NA61~\cite{Adhikary:2024epjc} \\
Xe+La  & ---$^a$ & \checkmark & NA61~\cite{Panova:2026} \\
Pb+Pb  & $|y| < 0.5$ & \checkmark & NA49~\cite{Afanasiev:2002mx,Alt:2007aa} \\
Au+Au  & $|y| < 0.1$ & AGS only & STAR~\cite{Adamczyk:2017iwn}, AGS~\cite{Alt:2007aa} \\
\end{tabular}
\end{ruledtabular}
{\footnotesize $^a$Only one preliminary mid-rapidity point exists
(150$A$~GeV, $y = 0.4$--$0.6$); not used.}
\end{table}

\begin{table*}
\caption{\label{tab:multiplicities}
$4\pi$ multiplicity data used in the per-participant analysis (all systems) and in the
statistical hadronization fits (Be+Be, Ar+Sc, Pb+Pb). $\avg{W}$: mean number of wounded
nucleons.}
\begin{ruledtabular}
\begin{tabular}{llll}
System (centrality) & $\avg{W}$ & Beam energies ($A$~GeV) & Hadrons \\
\hline
p+p (inel.) & 2 & 20, 31, 40, 80, 158 & $\pi^\pm, K^\pm, p, \bar p^{\,a}$ \\
Be+Be (0--20\%) & 7.99--8.15 & 19, 30, 40, 75, 150 & $\pi^\pm, K^\pm, \bar p^{\,b}$ \\
Ar+Sc (0--10\%) & 60.3--60.8 & 19, 30, 40, 75, 150 & $\pi^\pm, K^\pm, \bar p^{\,c}$ \\
Pb+Pb (0--7.2\%$^d$) & 349 (362$^d$) & 20, 30, 40, 80, 158 & $\pi^\pm, K^\pm, \phi, \Lambda, \bar\Lambda, \Xi^-, \bar\Xi^{+\,e}, \Omega^{\,f}$ \\
\end{tabular}
\end{ruledtabular}
{\footnotesize
\textsuperscript{a}\,$\bar p$ from 31$A$~GeV. \textsuperscript{b}\,$\bar p$ from 40$A$~GeV.
\textsuperscript{c}\,$\bar p$ from 30$A$~GeV. \textsuperscript{d}\,5\% at 158$A$~GeV.
\textsuperscript{e}\,$\bar\Xi^+$ from 30$A$~GeV. \textsuperscript{f}\,$\Omega+\bar\Omega$ at
40$A$, $\Omega$ and $\bar\Omega$ at 158$A$~GeV.}
\end{table*}

\section{Statistical hadronization with light- and strange-quark occupancies}
\label{sec:shm}

\subsection{Hadron resonance gas}
\label{sec:shm_hrg}

We use the statistical hadronization model with separate light- and strange-quark
phase-space occupancies, in the form developed by Letessier, Rafelski, and
collaborators, who treat $\gq$, entropy, and sudden hadronization together in
Ref.~\cite{Letessier:2002book} and implement the model in the SHARE
codes~\cite{Torrieri:2004zz,Torrieri:2006xi,Petran:2013dva}.
Hadron yields are computed in an ideal hadron resonance gas with the complete
PDG2025 hadron list distributed with Thermal-FIST~\cite{Vovchenko:2019pjl}
(444 states including antiparticles), exact Bose--Einstein and Fermi--Dirac
statistics, zero-width resonances, and all strong and electromagnetic decays.
Weakly decaying hadrons are kept stable, matching the experimental definitions;
measured $\Lambda$ yields include $\Sigma^0$. The primordial density of hadron $i$ is
\begin{equation}
n_i = \frac{g_i}{2\pi^2}\int_0^\infty
\frac{p^2\,dp}{\lambda_i^{-1}\exp\!\left(\sqrt{p^2+m_i^2}/T\right) \pm 1},
\label{eq:shm_density}
\end{equation}
with
\begin{equation}
\lambda_i = \gq^{\,N^q_i}\,\gs^{\,N^s_i}\,
\exp\!\left(\frac{B_i\muB + Q_i\mu_Q + S_i\mu_S}{T}\right),
\label{eq:shm_fugacity}
\end{equation}
where $N^s_i$ is the number of strange quarks and antiquarks in hadron $i$ and
$N^q_i = 2 - N^s_i$ (mesons) or $3 - N^s_i$ (baryons) the number of light ones,
following the Thermal-FIST convention~\cite{Vovchenko:2019pjl}. The phase-space
occupancies $\gq$ and $\gs$ measure the light- and strange-quark content at
hadronization relative to chemical equilibrium,
$\gq = \gs = 1$~\cite{Letessier:1998sz,Letessier:2000ay,Letessier:2002book}.
Values $\gq > 1$ describe an excess of light quark pairs, i.e., of entropy, over the
equilibrium hadron gas at the same temperature; $\gq$ is bounded by pion
Bose--Einstein condensation, $\gq^2\,e^{-\mu_Q/T} < e^{m_\pi/T}$ for $\pi^-$. Near this
bound the pion yields must be computed with the Bose--Einstein distribution, and the
pion entropy density is nearly twice its value at $\gq = 1$~\cite{Letessier:1999rm}.

The charge and strangeness chemical potentials follow from zero net strangeness and
$Q/B = Z/A$ of the colliding nuclei. For Pb+Pb all charges are treated
grand-canonically. For Be+Be and Ar+Sc strangeness is conserved exactly in a
correlation volume equal to the fireball volume (strangeness-canonical ensemble),
evaluated by Fourier projection of the strangeness generating function, while baryon
number and charge remain grand-canonical, as in Ref.~\cite{Becattini:2005xt} for
C+C and Si+Si. We checked the implementation against
Thermal-FIST~\cite{Vovchenko:2019pjl} in chemical equilibrium ($\gq = \gs = 1$) and the
grand-canonical ensemble, with $Q/B = 1/2$ and zero net strangeness. With the same
particle list (for this comparison, the PDG2025 list including light nuclei and
hypernuclei), it reproduces the Thermal-FIST yields of $\pi^+$ and $K^+$ after strong and
electromagnetic decays at 21 $(T,\muB)$ points covering the chemical freeze-out region,
from $T = 78$~MeV, $\muB = 758$~MeV to $T = 166$~MeV, $\muB = 24$~MeV, and those of
$\pi^-$ and $K^-$ at 12 of them, to better than 0.05\%, the precision of the reference
values. The fits themselves use the hadron list without nuclei.

\subsection{Fit procedure}
\label{sec:shm_fit}

The model is fitted to the $4\pi$ multiplicities of Table~\ref{tab:multiplicities} by
minimizing
\begin{equation}
\chi^2 = \sum_i \frac{\left(N_i^{\mathrm{exp}} - V n_i^{\mathrm{tot}}\right)^2}
{\sigma_{\mathrm{stat},i}^2 + \sigma_{\mathrm{sys},i}^2},
\label{eq:chi2_shm}
\end{equation}
where $n_i^{\mathrm{tot}}$ includes resonance feed-down. As in
Ref.~\cite{Becattini:2005xt}, $\avg{W}$ enters as a data point for the net baryon
number $V n_B$. The free parameters are $T$, $\muB$, $\gs$ and $V$, with $\gq$ either
fixed to unity (chemical equilibrium of light quarks) or scanned. Because $\gq$ and
$T$ are strongly correlated and the best fits lie close to the condensation boundary,
we quote $\chi^2$ profiles in $\gq$, obtained by refitting all other parameters at
each fixed $\gq$, rather than a best-fit value with a Gaussian error.

As a check, fitting with $\gq = 1$ the NA49 Pb+Pb data set used in
Ref.~\cite{Becattini:2005xt} reproduces the temperatures of their analysis~A to within
1~MeV at all five energies. Our $\muB$ is lower by 2--11~MeV and our $\gs$ by 2--7\%, by
up to about twice their quoted fit uncertainties, presumably because of the larger
resonance spectrum of PDG2025.

\subsection{What single ratios can and cannot determine}
\label{sec:shm_ratios}

Two properties of Eqs.~(\ref{eq:shm_density})--(\ref{eq:shm_fugacity}) govern the
interpretation of kaon-to-pion ratios. First, $K^+$ ($u\bar s$) carries $\gq\gs$ and
$\pi^+$ carries $\gq^2$, so $\avg{K^+}/\avg{\pi^+}$ depends on the occupancies only
through $\gs/\gq$: at a freeze-out point typical of 30$A$~GeV ($T = 139$~MeV,
$\muB = 425$~MeV), varying $\gq$ from 1 to 1.4 at fixed $\gs/\gq$ changes
$\avg{K^+}/\avg{\pi^+}$ by only 3\%. A strangeness
saturation factor extracted from $\avg{K^+}/\avg{\pi^+}$ alone with $\gq = 1$ is
therefore the ratio $\gs/\gq$, and $\gs/\gq \approx 1$ does not imply $\gs \approx 1$.
The non-equilibrium analyses quote $\gs/\gq$ alongside
$\gq$~\cite{Letessier:1999rm,Rafelski:2003wh}. The single ratios also depend on $\muB$.
Reference~\cite{Rafelski:2003wh} removed this dependence with the product of particle and
antiparticle ratios, $\sqrt{\avg{K^+}\avg{K^-}/(\avg{\pi^+}\avg{\pi^-})}$, in which the
fugacities cancel. The quotient of the two ratios is the complementary combination.
Second, for primordial kaons and pions the double ratio
$\mathcal{R} = (\avg{K^+}/\avg{\pi^+})/(\avg{K^-}/\avg{\pi^-})$ equals $e^{2\mu_S/T}$, so
the occupancies cancel explicitly (resonance decays lower $\mathcal{R}$ by up to about
25\% at SPS values of $\muB$), but $\mu_S$ is fixed by strangeness neutrality,
which balances kaons ($\propto\gq$) against hyperons ($\propto\gq^2$). As a result
$\mathcal{R}$ is nearly independent of $\gs$ (3.05--3.18 for $\gs = 0.3$--1.3 at the
same point) but rises with $\muB/T$ and with $\gq$ (2.83--3.80 for $\gq = 0.8$--1.4).
The double ratio is thus a probe of the baryon and light-quark sector of the fireball,
not of strangeness saturation.

\section{Entropy and strangeness per participant}
\label{sec:perpart}

\begin{figure*}
\includegraphics[width=\textwidth]{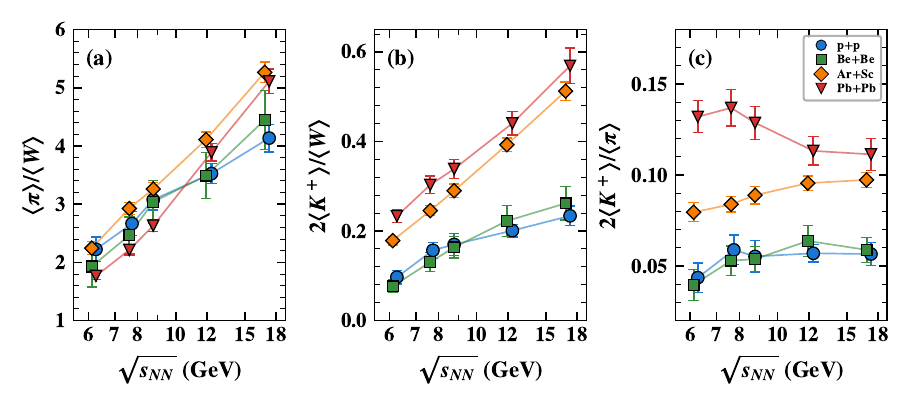}
\caption{\label{fig:perpart}
(a)~Pion yield per wounded nucleon, $\avg{\pi}/\avg{W}$ with
$\avg{\pi} = \frac{3}{2}(\avg{\pi^+} + \avg{\pi^-})$; (b)~strangeness per wounded nucleon,
$2\avg{K^+}/\avg{W}$; (c)~their ratio $2\avg{K^+}/\avg{\pi}$, as a function of
$\sqrts$ for inelastic p+p, central Be+Be, Ar+Sc (NA61/SHINE) and Pb+Pb (NA49), all in
$4\pi$ (Table~\ref{tab:multiplicities}). Error bars combine statistical and systematic
uncertainties of the yields and of $\avg{W}$; lines guide the eye.}
\end{figure*}

The horn mechanism described in Sec.~\ref{sec:intro} makes a statement about two
separate yields: strangeness per participant and entropy per participant. We use the
total pion yield $\avg{\pi} = \frac{3}{2}(\avg{\pi^+} + \avg{\pi^-})$ per wounded
nucleon as the entropy measure and $2\avg{K^+}/\avg{W}$ as the strangeness measure:
in baryon-rich matter nearly all $\bar s$ quarks are carried by $K^+$ and $K^0$, and
$\avg{K^0} \approx \avg{K^+}$ by isospin. Their ratio,
$2\avg{K^+}/\avg{\pi}$, is the kaon-to-pion ratio up to the isospin factor.
Using the particle multiplicity per baryon as a measure of the entropy content goes back to
Ref.~\cite{Letessier:1992xd}, which found in S-induced reactions at 200$A$~GeV more than
twice the entropy per baryon that an equilibrium hadron gas fitted to the particle ratios
provides. Strangeness per baryon and per entropy were followed as functions of collision
energy in Ref.~\cite{Rafelski:2003wh}, where both are obtained from statistical
hadronization fits of the hadron yields. Here we use the measured yields per wounded nucleon directly; this
requires no fit and allows a comparison across systems.
Figure~\ref{fig:perpart} shows the three quantities for the four systems with
published $4\pi$ data.

Three features stand out. First, strangeness per participant jumps between Be+Be and
Ar+Sc at every energy, by a factor 1.8--2.3, while the pion yield per participant
changes by only 7--19\% [Figs.~\ref{fig:perpart}(a),(b)]. The system-size step in
$\avg{K^+}/\avg{\pi^+}$ (the ``onset of fireball'') is therefore a step in strangeness
production, as expected if strangeness production is amplified by multiple collisions
per participant. Second, the horn is present only in Pb+Pb [Fig.~\ref{fig:perpart}(c)]:
between 30$A$ and 158$A$~GeV, $2\avg{K^+}/\avg{\pi}$ changes by at most 16\% for p+p,
Be+Be and Ar+Sc (rising slowly for Ar+Sc), whereas it falls by 19\% for Pb+Pb. Third, the origin of the horn is on
the entropy side. Pb+Pb has the lowest pion yield per participant of all four systems at
20$A$--40$A$~GeV (2.21 at 30$A$~GeV, against 2.67 for p+p and 2.93 for Ar+Sc), and it
catches up by 80$A$--158$A$~GeV, the known kink~\cite{Gazdzicki:2010iv}. Between 30$A$
and 158$A$~GeV strangeness per participant grows at a similar rate in Pb+Pb (factor 1.9)
and Ar+Sc (2.1), but the pion yield per participant grows by a factor 2.3 in Pb+Pb
against 1.8 in Ar+Sc. The drop of $\avg{K^+}/\avg{\pi^+}$ above 30$A$~GeV is thus an
entropy effect specific to the largest system.

The $\avg{W}$ values of NA49 (Glauber) and NA61/SHINE (EPOS with the experimental
event selection) are not defined identically, and the Be+Be $\avg{W}$ is uncertain by
about 10\%. These differences are small compared with the 32\% difference between the
Ar+Sc and Pb+Pb pion yields per participant at 30$A$~GeV.

\section{Equilibrium and non-equilibrium fits}
\label{sec:fits}

\subsection{Central Pb+Pb}
\label{sec:fits_pbpb}

\begin{figure*}
\includegraphics[width=\textwidth]{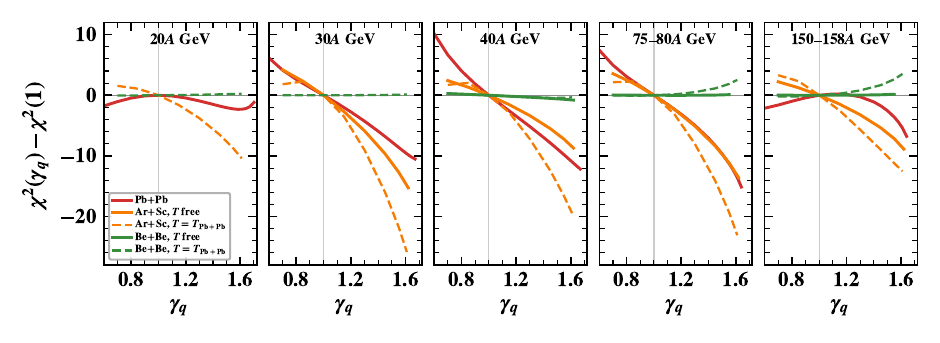}
\caption{\label{fig:gq_profiles}
Profiles $\chi^2(\gq) - \chi^2(\gq = 1)$ of the statistical hadronization fits, with
$T$, $\muB$, $\gs$ and $V$ refitted at each fixed $\gq$, for central Pb+Pb (NA49),
Ar+Sc and Be+Be (NA61/SHINE), grouped by beam energy. Solid lines: $T$ free (for
Be+Be and Ar+Sc only where $\bar p$ is measured); dashed lines: $T$ fixed at each $\gq$
to the Pb+Pb best-fit value at the nearest beam energy. Curves end where pions
condense.}
\end{figure*}

Table~\ref{tab:shm_fits} and Fig.~\ref{fig:gq_profiles} summarize the fits. For central
Pb+Pb with $\gq = 1$, the fitted temperatures (140--151~MeV) and baryon chemical
potentials (472--245~MeV) follow the chemical freeze-out systematics, with
$\gs = 0.68$--0.81, but the fit quality is poor from 30$A$~GeV upward
($\chindf = 14.2/6$--$25.0/6$). The fits fall short in the pion yields (pulls $-0.8$ to
$-1.8$) and in the $\Lambda$ and $\Xi$ yields together (down to $-2.2$), and they do
not reproduce the $K^+/K^-$ asymmetry (the individual yields and pulls are listed in
Appendix~\ref{app:fit_tables}). With $\gq$ free, $\chi^2$ decreases as $\gq$
approaches the pion condensation limit and reaches its minimum at, or just below, that
limit ($\gq = 1.64$--1.67), with $T = 129$--136~MeV and
$\gs = 1.4$--1.7; the improvement is $\Delta\chi^2 = -10.5$, $-12.2$, $-15.1$ and $-6.8$
at 30$A$, 40$A$, 80$A$ and 158$A$~GeV. The light-quark excess restores the pion and
hyperon yields and removes part of the $K^+/K^-$ tension. At 20$A$~GeV the profile is
flat on both sides of $\gq = 1$ ($|\Delta\chi^2| \le 2.3$): the data do not
constrain $\gq$ there.

Because the minima lie at the physical boundary of the parameter space, Wilks' theorem
does not apply and we do not convert $\Delta\chi^2$ into a significance. The pattern,
however, is clear: chemical equilibrium of light quarks is disfavored from the horn
energy upward, most robustly at 30$A$--80$A$~GeV (Sec.~\ref{sec:fits_robust}), with the
light-quark phase space saturated at the condensation limit, while below it the data are
compatible with either. With the final NA49 data and an independent implementation,
this confirms the result of Letessier, Rafelski, and
collaborators~\cite{Letessier:2008,Rafelski:2008av}, who found $\gq$ near
$e^{m_\pi/2T}$ from 30$A$~GeV upward. Their under-saturated $\gq$ at 20$A$~GeV, and
hence the change of hadronization dynamics between 20$A$ and 30$A$~GeV, is neither
confirmed nor excluded by our fits, which do not constrain $\gq$ at 20$A$~GeV. At $\gq$ near its limit, $\gs/\gq \approx
0.9$--1.0 in Pb+Pb: the value of $\gs$ close to unity found by fits with $\gq = 1$
reflects $\gs/\gq$, not strangeness chemical equilibrium.

\begin{table*}
\caption{\label{tab:shm_fits}
Statistical hadronization fits of $4\pi$ multiplicities. Left: chemical equilibrium of
light quarks ($\gq = 1$). Right: minimum of the $\chi^2$ profile in $\gq$ (with one
parameter more, ndf reduced by one). Pb+Pb: grand-canonical; Be+Be and Ar+Sc:
strangeness-canonical with correlation volume equal to $V$. $T$ and $\muB$ in MeV.
Fits of Be+Be and Ar+Sc without $\bar p$ (19$A$~GeV, and 30$A$~GeV for Be+Be) do not
constrain $T$ and are not listed. $^\dagger$Minimum at the edge of the scanned range,
close to the pion condensation limit. $^\ddagger$Flat profile; $\gq$ not constrained.}
\begin{ruledtabular}
\begin{tabular}{lc|cccc|cccccc}
 & & \multicolumn{4}{c|}{$\gq = 1$} & \multicolumn{6}{c}{$\gq$ free} \\
System & $E_{\mathrm{beam}}$ ($A$~GeV) & $T$ & $\muB$ & $\gs$ & $\chindf$ & $\gq$ & $T$ & $\muB$ & $\gs$ & $\chindf$ & $\Delta\chi^2$ \\
\hline
Pb+Pb & 20  & 140 & 472 & 0.77 & 11.3/5 & 1.58$^\ddagger$ & 126 & 426 & 1.41 & 9.0/4 & $-2.3$ \\
      & 30  & 144 & 417 & 0.80 & 14.2/6 & 1.67$^\dagger$ & 129 & 379 & 1.67 & 3.6/5 & $-10.5$ \\
      & 40  & 146 & 375 & 0.75 & 19.7/7 & 1.67$^\dagger$ & 131 & 345 & 1.57 & 7.5/6 & $-12.2$ \\
      & 80  & 151 & 291 & 0.68 & 25.0/6 & 1.64$^\dagger$ & 136 & 270 & 1.42 & 9.9/5 & $-15.1$ \\
      & 158 & 151 & 245 & 0.81 & 21.3/8 & 1.64$^\dagger$ & 136 & 228 & 1.66 & 14.5/7 & $-6.8$ \\
\hline
Ar+Sc & 30  & 134 & 373 & 0.48 & 28.7/2 & 1.62$^\dagger$ & 125 & 346 & 1.01 & 13.4/1 & $-15.3$ \\
      & 40  & 137 & 342 & 0.52 & 14.8/2 & 1.62$^\dagger$ & 128 & 317 & 1.08 & 6.1/1 & $-8.7$ \\
      & 75  & 143 & 292 & 0.57 & 21.0/2 & 1.62$^\dagger$ & 133 & 268 & 1.19 & 7.4/1 & $-13.6$ \\
      & 150 & 147 & 239 & 0.60 & 14.2/2 & 1.62$^\dagger$ & 137 & 219 & 1.25 & 5.3/1 & $-8.9$ \\
\hline
Be+Be & 40  & 145 & 346 & 0.38 & 1.2/2 & --$^\ddagger$ & & & & & $-0.8$ \\
      & 75  & 156 & 297 & 0.44 & 0.1/2 & --$^\ddagger$ & & & & & $0.0$ \\
      & 150 & 155 & 253 & 0.41 & 0.2/2 & --$^\ddagger$ & & & & & $0.0$ \\
\end{tabular}
\end{ruledtabular}
\end{table*}

\subsection{Ar+Sc and Be+Be}
\label{sec:fits_small}

The fits of the Be+Be~\cite{Acharya:2021ljw} and Ar+Sc~\cite{Adhikary:2024epjc}
multiplicities with $\gq$ free are the main new result of this work. The Ar+Sc fits also disfavor $\gq = 1$: $\chi^2$ drops by 8.7--15.3 units between
$\gq = 1$ and the edge of the scanned range ($\gq = 1.62$) at 30$A$--150$A$~GeV, and by
10--26 units when $T$ is tied to the Pb+Pb values, which also includes 19$A$~GeV
(Fig.~\ref{fig:gq_profiles}). The reason, however, differs from Pb+Pb. With $T$ free the
pion yields of Ar+Sc are described at $\gq = 1$; the entire tension lies in the kaons,
with $K^+$ underestimated and $K^-$ overestimated (pulls $-4.3$ and $+2.6$ at 30$A$~GeV).
The fit therefore does not find an entropy excess in Ar+Sc. It finds a $K^+/K^-$
asymmetry larger than a single thermal source can produce with the baryon chemical
potential fixed by $\avg{W}$ and $\avg{\bar p}$; $\gq > 1$ reduces this tension only
indirectly, by raising $\mu_S$ through strangeness neutrality
(Sec.~\ref{sec:shm_ratios}), and the remaining $\chi^2$ (5.3--13.4 for one degree of
freedom) shows that no single hadronizing source describes Ar+Sc. We return to this in
Secs.~\ref{sec:results_doubleratio} and~\ref{sec:discussion}.

For Be+Be, where only pions, kaons and (from 40$A$~GeV) antiprotons are measured and
the pion uncertainties are 8--12\% (about 20\% at 19$A$~GeV), $\chi^2$ never exceeds 1.5 (with $T$ free, two
degrees of freedom) or 4.2 (with $T$ tied to Pb+Pb, three degrees of freedom) for $\gq$
between 0.7 and the condensation limit; with $T$ tied, $\gq$ near the limit is mildly
disfavored at 75$A$--150$A$~GeV ($\Delta\chi^2 \le 3.5$). The data are compatible with
chemical equilibrium of light quarks and do not require $\gq > 1$. The quantity they
determine is $\gs/\gq \approx 0.4$, well below the Ar+Sc values
($\gs/\gq = 0.48$--0.60 at $\gq = 1$, 0.62--0.77 at the best $\gq$). Since exact
strangeness conservation is included, this difference is not a canonical-suppression
effect, as long as the correlation volume is the fireball volume. p+p is not fitted:
it requires exact conservation of baryon number and charge as well, beyond the
ensemble used here.

\subsection{What the preference for $\gq > 1$ rests on}
\label{sec:fits_robust}

The preference for $\gq > 1$ rests on differences of 7--15 units of $\chi^2$, with minima
at the boundary of the parameter space. To identify what carries it, we repeated the
$\chi^2$ profiles with one element of the analysis changed at a time
(Table~\ref{tab:robustness}). The elements varied follow from earlier work. An entropy
excess shows up as more produced particles per baryon than an equilibrium hadron gas
provides~\cite{Letessier:1992xd}. In sudden hadronization it is carried by $\gq$ near the
condensation limit, where the pion yields are Bose--Einstein
enhanced~\cite{Letessier:1999rm}. The stability of non-equilibrium fits against the pion
yield was tested in Ref.~\cite{Rafelski:2003wh} by scaling it by 0.8--1.2. Here we
quantify how much of $\Delta\chi^2$ each of these ingredients carries.

The result does not depend on the resonance spectrum or on the data compilation.
Truncating the hadron list at 1.8~GeV (255 of the 444 states) changes $\Delta\chi^2$ by at
most two units, and the Pb+Pb data set of Ref.~\cite{Becattini:2005xt} gives
$\Delta\chi^2 = -8.8$ to $-11.0$ from 30$A$~GeV upward. At 30$A$--80$A$~GeV no single
yield carries the preference: removing any one hadron yield, or both pion yields, leaves
$\Delta\chi^2 \le -3.9$, the weakest case being the removal of $K^-$ at 30$A$~GeV.

Two ingredients are essential. The first is the net-baryon number. Without $\avg{W}$, the
equilibrium fits at 30$A$--80$A$~GeV are acceptable ($\chi^2 = 5.8/5$, $6.3/6$ and
$6.1/5$ at $\gq = 1$), and $\Delta\chi^2$ shrinks to between $-2.7$ and $-3.4$. The yields
of the produced hadrons are thus compatible with chemical equilibrium among themselves;
it is their abundance relative to the net-baryon number that requires $\gq > 1$. The
second is quantum statistics. With Boltzmann statistics for all hadrons, $\Delta\chi^2$ is
only $-0.7$ to $-2.6$ at 30$A$--80$A$~GeV. If the pion yields are removed as well, Boltzmann
and quantum statistics give nearly the same result ($\Delta\chi^2 = -4.9$ and $-5.6$ at
30$A$~GeV), so the difference comes entirely from the pions. The strange hadrons per net
baryon favor $\gq > 1$ irrespective of statistics; the pions do so only through their
Bose--Einstein enhancement near $\gq = e^{m_\pi/2T}$, where low-momentum pion states are
highly occupied, as anticipated in Ref.~\cite{Letessier:1999rm}. The Boltzmann approximation is not a valid description in this region, so
this comparison identifies the mechanism and is not a systematic uncertainty. In the
language of Sec.~\ref{sec:intro}, the Pb+Pb yields at 30$A$--80$A$~GeV contain more
produced hadrons, and in particular more pions, i.e., more entropy, per net baryon than a
hadron gas in chemical equilibrium can accommodate, as found earlier for S-induced
reactions~\cite{Letessier:1992xd}. The non-equilibrium fits store this excess in
over-saturated light-quark phase
space~\cite{Letessier:1999rm,Rafelski:2000by,Letessier:2008}.

At 158$A$~GeV the preference is fragile. It disappears without the pion yields
($\chi^2 = 4.2/6$ at $\gq = 1$, and no preference remains) or without $K^-$
($\Delta\chi^2 = -0.6$), and removing $\bar\Lambda$ reduces it to $-2.6$. The profile also falls toward $\gq < 1$ there
($\chi^2 = 19.2$ at $\gq = 0.6$, against 21.3 at $\gq = 1$ and 14.5 at the minimum), and
without $\avg{W}$ or with Boltzmann statistics this branch becomes the lower one. At
20$A$~GeV no variation gives $|\Delta\chi^2| > 4.2$, on either side of $\gq = 1$.

For Ar+Sc the variations confirm the reading of Sec.~\ref{sec:fits_small}. Removing one of
the pion yields changes $\Delta\chi^2$ by at most 1.2 units (both cannot be removed without
losing all degrees of freedom), whereas removing any one of $\avg{W}$, $K^+$, $K^-$ and
$\bar p$ removes the preference ($|\Delta\chi^2| \le 1.1$). The Ar+Sc result is a conflict between the $K^+/K^-$ ratio
and the baryon chemical potential fixed by $\avg{W}$ and $\bar p$, not an excess of pions.

\begin{table*}
\caption{\label{tab:robustness}
Robustness of the preference for $\gq > 1$: $\Delta\chi^2 = \chi^2_{\min} - \chi^2(\gq = 1)$
with one element of the analysis changed at a time; the first row is the baseline of
Table~\ref{tab:shm_fits}. For the removal of $\phi$ or one hyperon, the weakest (least
negative) value is given. \textsuperscript{a}\,Minimum at $\gq < 1$. --: not applicable, or
no degrees of freedom left.}
\begin{ruledtabular}
\begin{tabular}{l|ccccc|cccc}
 & \multicolumn{5}{c|}{Pb+Pb} & \multicolumn{4}{c}{Ar+Sc} \\
$E_{\mathrm{beam}}$ ($A$~GeV) & 20 & 30 & 40 & 80 & 158 & 30 & 40 & 75 & 150 \\
\hline
Baseline (Table~\ref{tab:shm_fits}) & $-2.3$ & $-10.5$ & $-12.2$ & $-15.1$ & $-6.8$ & $-15.3$ & $-8.7$ & $-13.6$ & $-8.9$ \\
Hadron list with $m < 1.8$~GeV & $-2.0$\textsuperscript{a} & $-9.8$ & $-11.2$ & $-13.2$ & $-4.9$ & $-15.5$ & $-8.8$ & $-13.5$ & $-8.7$ \\
Pb+Pb data of Ref.~\cite{Becattini:2005xt} & $-1.4$\textsuperscript{a} & $-10.5$ & $-10.3$ & $-8.8$ & $-11.0$ & -- & -- & -- & -- \\
Without $\pi^+$ and $\pi^-$ & $-0.7$\textsuperscript{a} & $-5.6$ & $-8.0$ & $-9.2$ & $0.0$ & -- & -- & -- & -- \\
Without $K^+$ & $-1.8$\textsuperscript{a} & $-6.1$ & $-8.3$ & $-10.4$ & $-6.8$ & $0.0$ & $0.0$ & $0.0$ & $0.0$ \\
Without $K^-$ & $-0.8$\textsuperscript{a} & $-3.9$ & $-5.7$ & $-8.0$ & $-0.6$ & $0.0$ & $0.0$ & $0.0$ & $0.0$ \\
Without $\phi$ or one hyperon & $-1.7$ & $-9.2$ & $-9.1$ & $-10.9$ & $-2.6$ & -- & -- & -- & -- \\
Without $\bar p$ & -- & -- & -- & -- & -- & $-1.1$ & $0.0$ & $0.0$ & $0.0$ \\
Without $\avg{W}$ & $-1.1$\textsuperscript{a} & $-2.7$ & $-3.2$ & $-3.4$ & $-5.0$\textsuperscript{a} & $0.0$ & $0.0$ & $0.0$ & $0.0$ \\
Boltzmann statistics & $-4.1$\textsuperscript{a} & $-1.5$ & $-2.6$ & $-0.7$ & $-5.1$\textsuperscript{a} & $-3.2$ & $-1.9$ & $-2.8$ & $-1.7$ \\
\end{tabular}
\end{ruledtabular}
\end{table*}

\section{Models along the system-size axis}
\label{sec:models}

\subsection{Models and $\chi^2$ definition}
\label{sec:models_def}

\paragraph*{SMASH.}
SMASH (Simulating Many Accelerated Strongly-interacting
Hadrons)~\cite{Weil:2016zrk,SMASH:2023zenodo} is a hadronic transport model that
propagates hadrons and resonances via binary scattering, resonance
formation, and string excitation~\cite{Mohs:2019iee}, without any partonic phase.
Version 3.3 is used with the default particle list and collision settings, for all six
systems at 5--10 beam energies each, with about $5\times 10^4$ or more events per point. For
nucleus--nucleus collisions the 0--5\% most central events are selected by impact
parameter. This is more central than the Be+Be (0--20\%), Ar+Sc and Xe+La (0--10\%) and
Pb+Pb (0--7.2\%) data; since the kaon-to-pion ratios vary slowly with centrality in
central collisions the mismatch is small compared with the model--data differences
below, but it is not corrected for.

\paragraph*{PHSD.}
The Parton-Hadron-String Dynamics approach is used as implemented in the PHSD--PHQMD
code v.1.0 in PHSD mode~\cite{Cassing:2008sv,Cassing:2009vt,Moreau:2019vhw}. It
describes the partonic phase by dynamical quasiparticles with lattice-QCD
thermodynamics, hadronizes by covariant transition rates, and includes chiral symmetry
restoration in the string decay via the Schwinger mechanism~\cite{Cassing:2016,Palmese:2016rtq},
which in this code is active for nucleus--nucleus collisions at beam kinetic energies up
to 250$A$~GeV and inactive for p+p. The calculations cover all systems of this work.
Centrality is selected as in the experiments, from the event-by-event charged-particle
multiplicity in full phase space, taking the most central fraction of events that
corresponds to each experimental class. The finite sampling of impact parameters introduces an additional systematic uncertainty in the absolute yields that is not included in the quoted statistical errors, although the kaon-to-pion ratios are much less sensitive to this effect. Au+Au at 27 and 39~GeV
is not used, because the selected samples there are not central ($\avg{b} = 4.2$ and
5.4~fm). Mid-rapidity yields are taken in $|y| < 0.5$ in the center-of-mass frame.

\paragraph*{Glauber core--corona.}
The core--corona model~\cite{Werner:2007,Becattini:2008yn,Aichelin:2008mi} splits each
collision into a core, described by a grand-canonical hadron resonance gas in chemical
equilibrium, and a corona, described by SMASH:
\begin{equation}
\label{eq:cc}
\left(\frac{K^+}{\pi^+}\right)_{\mathrm{CC}} =
\fcore \left(\frac{K^+}{\pi^+}\right)_{\mathrm{HRG}}
+ (1 - \fcore) \left(\frac{K^+}{\pi^+}\right)_{\mathrm{SMASH}} .
\end{equation}
The core values are the Thermal-FIST values of Sec.~\ref{sec:shm_hrg}, interpolated in
$\sqrts$; their $(T,\muB)$ follow the freeze-out curve of Ref.~\cite{Cleymans:2005xv} to
within 1.5~MeV in $T$ above $\sqrts = 4$~GeV. Outside their range (Au+Au at 2.35, 27 and
39~GeV) the core is computed on that curve with our implementation and the same particle
list. The core fraction $\fcore$ is the fraction of participants with at least
$n_{\mathrm{coll}}^{\mathrm{thr}}$ binary collisions in a Glauber
model~\cite{Miller:2007ri,Loizides:2017ack} at the experimental centrality. The
threshold $n_{\mathrm{coll}}^{\mathrm{thr}} = 1.21$ is calibrated on the Pb+Pb $4\pi$
$\avg{K^+}/\avg{\pi^+}$ (5 points, one parameter) and transferred unchanged to all
other systems. At 30$A$~GeV this gives $\fcore = 0$ (p+p), 0.21 (Be+Be), 0.58 (Ar+Sc),
0.71 (Xe+La) and 0.74 (Pb+Pb). Mixing at the level of ratios, Eq.~(\ref{eq:cc}),
assumes equal pion yields per participant in core and corona. The core is in chemical
equilibrium by construction;
Sec.~\ref{sec:fits} shows that this is an idealization, and we use the model as a
geometric description of the corona fraction. An energy-dependent threshold,
$n_{\mathrm{coll}}^{\mathrm{thr}} = t_0 + t_1\ln\sqrts$, improves the Pb+Pb description
($\Delta\chi^2 = 6.0$ for one additional parameter; bootstrap $p = 0.010$), but we keep
the constant threshold.

\paragraph*{SMES.}
The Statistical Model of the Early Stage~\cite{Gazdzicki:1998vd} is used in two
variants. The fitted variant has three parameters (the critical temperature, the
exponent of the energy-density mapping, and the onset energy) adjusted to the NA49
Pb+Pb and AGS Au+Au mid-rapidity data~\cite{Ahle:2000wq,Klay:2003zf} ($T_c = 165$~MeV,
$\sqrts_{\mathrm{onset}} = 7.62$~GeV); STAR Au+Au and all smaller systems are
predictions. The canonical variant (SMES--CE) has no fitted parameters: it takes the
volume parametrization, $T_c$ and bag constant from Ref.~\cite{Poberezhnyuk:2015} and
introduces system-size dependence only through exact strangeness conservation, with the
mixed phase between $\sqrts = 7.36$ and 11.75~GeV as an output. Both variants describe
the onset through a mixed phase.

\paragraph*{$\chi^2$ definition.}
All model--data comparisons use
$\chi^2 = \sum_i (d_i - m_i)^2/(\sigma_{d,i}^2 + \sigma_{m,i}^2)$, with the experimental
statistical and systematic uncertainties added in quadrature and the model statistical
uncertainty $\sigma_m$. Model and data are matched energy by energy. We quote
$\chi^2/\mathrm{ndf}$ with $\mathrm{ndf} = n - k$, where $k$ counts parameters fitted
to that system; for predictions ($k = 0$) this is the mean squared pull.

\subsection{Transport baselines}
\label{sec:results_transport}

\begin{figure*}
\includegraphics[width=\textwidth]{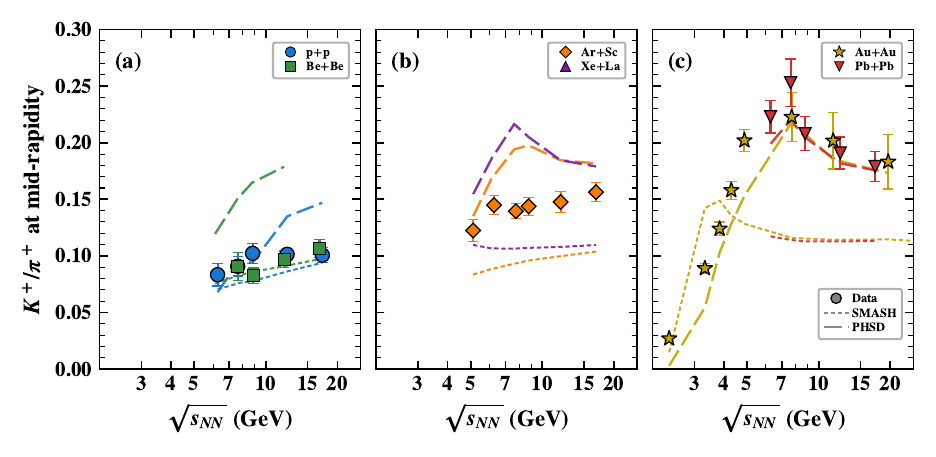}
\caption{\label{fig:baseline_midrap}
Mid-rapidity $K^+/\pi^+$ excitation functions.
(a)~Light systems: p+p and Be+Be;
(b)~intermediate systems: Ar+Sc and Xe+La (no mid-rapidity data for Xe+La);
(c)~heavy systems: Au+Au (AGS and STAR) and Pb+Pb.
Symbols: data; short-dashed lines: SMASH-3.3; long-dashed lines: PHSD--PHQMD v.1.0 in
PHSD mode. PHSD produces a horn in Pb+Pb and Au+Au, but also in Ar+Sc and Xe+La.}
\end{figure*}

\begin{figure}
\includegraphics[width=\columnwidth]{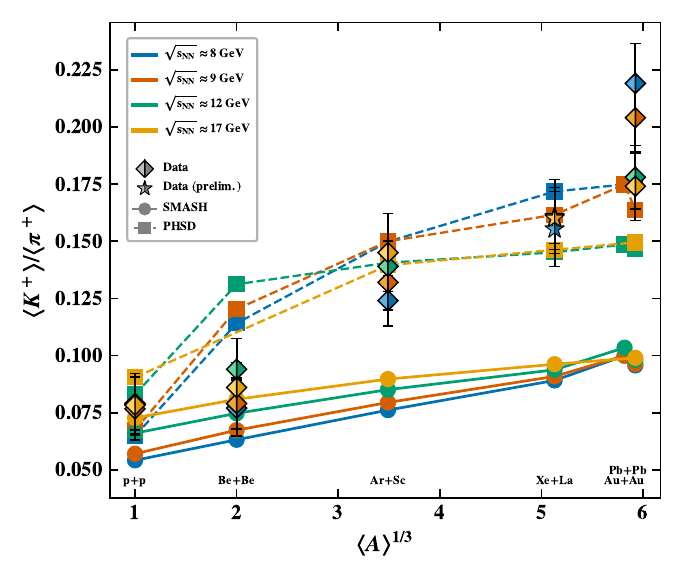}
\caption{\label{fig:system_size}
System-size dependence of $\avg{K^+}/\avg{\pi^+}$ ($4\pi$) at selected SPS energies,
as a function of $\avg{A}^{1/3}$. Diamonds: data (stars: preliminary Xe+La); circles
and solid lines: SMASH; squares and dashed lines: PHSD. Colors distinguish the
energies.}
\end{figure}

Figure~\ref{fig:baseline_midrap} compares the transport calculations with the
mid-rapidity data. SMASH produces no maximum of $K^+/\pi^+$ at SPS energies: the ratio
rises with $\sqrts$ for p+p, Be+Be and Ar+Sc and is nearly flat for Xe+La and Pb+Pb (for
Au+Au it peaks at AGS energies, near $\sqrts = 3.8$~GeV). It is close to the Be+Be data
($\chindf = 0.70$), somewhat below p+p (4.45), and underpredicts Ar+Sc, Pb+Pb and STAR
Au+Au by factors of 1.5--2.2 ($\chindf = 34.7$ for Ar+Sc and 38.6 for Pb+Pb,
Table~\ref{tab:model_chi2}).
PHSD, with partonic degrees of freedom and chiral symmetry restoration, produces a horn:
in central Pb+Pb its mid-rapidity $K^+/\pi^+$ rises from 0.199 at 20$A$ to 0.220 at
30$A$~GeV and falls to 0.176 at 158$A$~GeV, and in Au+Au it peaks at
$\sqrts = 7.7$~GeV. It describes the Pb+Pb ($\chindf = 1.15$) and STAR Au+Au (0.22)
mid-rapidity data, although the Pb+Pb peak height is underestimated (0.220 against
0.253). PHSD, however, also produces a horn in Ar+Sc and Xe+La
[Fig.~\ref{fig:baseline_midrap}(b)], with a maximum at 30$A$--40$A$~GeV that lies 9\% and
21\% above the 150$A$~GeV value (25\% in Pb+Pb), whereas the Ar+Sc data show no such
maximum; at mid-rapidity PHSD overpredicts Ar+Sc by up to 40\%
($\chindf = 22.4$) and Be+Be by up to a factor of two (72.8). In $4\pi$, where the experimental
uncertainties are larger, PHSD describes Ar+Sc and the preliminary Xe+La data
($\chindf = 1.6$ and 1.0), but its Xe+La excitation function has a maximum at
30$A$~GeV 17\% above its 150$A$~GeV value, while the preliminary data are flat within
4\%. The system-size dependence at fixed energy (Fig.~\ref{fig:system_size}) summarizes
the pattern: SMASH scales smoothly and stays far below the data for all systems larger
than Be+Be, while PHSD is close to the $4\pi$ data for p+p, Ar+Sc and Xe+La but too high
for Be+Be and too low for Pb+Pb near the horn energy.

\subsection{Core--corona and SMES}
\label{sec:results_cc}

\begin{figure*}
\includegraphics[width=\textwidth]{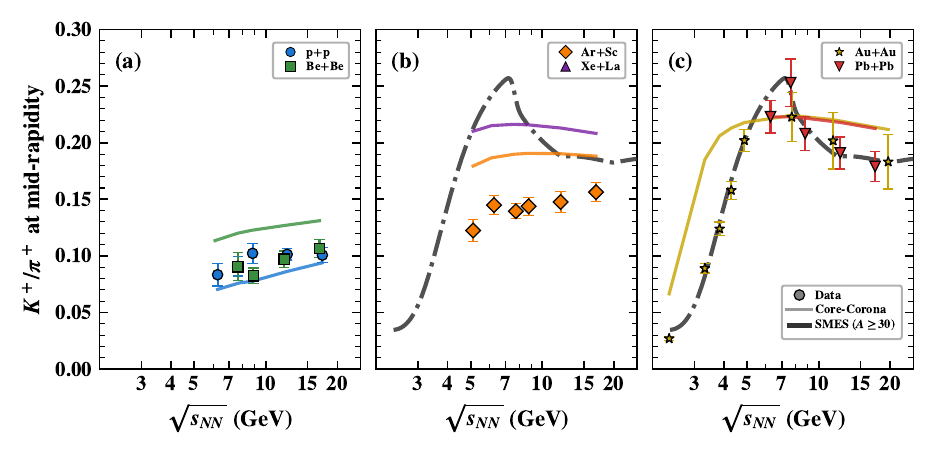}
\caption{\label{fig:model_comparison}
Mid-rapidity $K^+/\pi^+$ compared with the core--corona model (solid lines, Glauber core
fraction at the experimental centrality) and the fitted SMES (dash-dotted; shown for
$A \geq 30$). Panels as in Fig.~\ref{fig:baseline_midrap}.}
\end{figure*}

Table~\ref{tab:model_chi2} collects the $\chi^2$ values of all models for
$\avg{K^+}/\avg{\pi^+}$ in both acceptances, and Fig.~\ref{fig:model_comparison} shows
the core--corona and SMES curves at mid-rapidity. With its single parameter calibrated on
Pb+Pb ($\chindf = 1.69$ in $4\pi$), the core--corona model improves Ar+Sc
($\chindf$: $14.9 \to 9.3$ in $4\pi$) and the preliminary Xe+La ($23.3 \to 5.4$) relative
to SMASH, and it describes the STAR Au+Au data ($\chindf = 0.89$). It overshoots Be+Be,
where SMASH alone is adequate, and it fails for the AGS Au+Au data
($\sqrts < 5$~GeV), where the equilibrium core overpredicts $K^+/\pi^+$ by up to a factor
1.6 in $4\pi$ and 2.5 at mid-rapidity at the lowest energies. The step in $\fcore$ from 0.21 (Be+Be) to 0.58 (Ar+Sc) follows from Glauber
geometry alone: Ar+Sc produces enough binary collisions per participant to exceed the
threshold, Be+Be does not. It is therefore a geometric counterpart of the strangeness
step of Sec.~\ref{sec:perpart}, but the model does not reproduce the Ar+Sc energy
dependence at mid-rapidity ($\chindf = 30.7$).

The fitted SMES reproduces the Pb+Pb and Au+Au horn with the onset at
$\sqrts = 7.62$~GeV, and SMES--CE, without fitted parameters, predicts the mixed-phase
onset at 7.36~GeV and describes Pb+Pb and STAR Au+Au ($\chindf = 2.46$ and 0.67). Both
overpredict Ar+Sc strongly ($\chindf = 87$ and 132 at mid-rapidity), and
SMES--CE also overpredicts Be+Be. In SMES--CE the energy density does not depend on the
system size, which enters only through canonical suppression; this makes Be+Be approach
Pb+Pb, whereas the data show Be+Be tracking p+p. Both variants rely on a mixed phase,
which is subject to the interferometry constraint discussed in Sec.~\ref{sec:intro}.

\begin{table*}
\caption{\label{tab:model_chi2}
$\chindf$ of model predictions for $K^+/\pi^+$, per system and acceptance
[$\chi^2$ definition in Sec.~\ref{sec:models_def}; ndf $= n$ except CC for Pb+Pb $4\pi$
(calibration, ndf $= n - 1$)]. $n$: number of data points (for PHSD Be+Be: 3 at
mid-rapidity and 4 in $4\pi$; PHSD STAR Au+Au: 3; SMES and SMES--CE AGS Au+Au: 4). Au+Au is split into AGS
($\sqrts < 5$~GeV) and STAR ($\sqrts = 7.7$--39~GeV). Xe+La uses preliminary
data~\cite{Panova:2026}.}
\begin{ruledtabular}
\begin{tabular}{llc|ccccc}
Acceptance & System & $n$ & SMASH & PHSD & CC & SMES$^a$ & SMES--CE \\
\hline
$4\pi$ & p+p   & 5 & 1.77 & 0.68 & 1.77$^b$ & -- & -- \\
       & Be+Be & 5 & 0.94 & 9.04 & 4.16 & -- & -- \\
       & Ar+Sc & 6 & 14.9 & 1.57 & 9.32 & 39.5 & -- \\
       & Xe+La (prelim.) & 4 & 23.3 & 0.98 & 5.38 & 6.41 & -- \\
       & Pb+Pb & 5 & 42.0 & 6.06 & 1.69$^c$ & 0.22$^a$ & -- \\
       & Au+Au (AGS) & 5 & 28.0 & 56.9 & 34.9 & 5.66$^a$ & -- \\
\hline
Mid-rapidity & p+p   & 5 & 4.45 & 7.49 & 4.45$^b$ & -- & 9.34 \\
       & Be+Be & 4 & 0.70 & 72.8 & 16.6 & -- & 160 \\
       & Ar+Sc & 6 & 34.7 & 22.4 & 30.7 & 86.8 & 132 \\
       & Pb+Pb & 5 & 38.6 & 1.15 & 2.60 & 0.73$^a$ & 2.46 \\
       & Au+Au (STAR) & 5 & 12.4 & 0.22 & 0.89 & 0.19 & 0.67 \\
       & Au+Au (AGS) & 5 & 66.2 & 76.4 & 320 & 1.71$^a$ & 190 \\
\end{tabular}
\end{ruledtabular}
{\footnotesize
$^a$Three parameters fitted to the NA49 Pb+Pb and AGS Au+Au mid-rapidity data; these rows
are therefore not predictions. \textsuperscript{b}\,$\fcore = 0$: identical to SMASH.
$^c$In-sample (calibration of the threshold).}
\end{table*}

\subsection{Xe+La: preliminary comparison and test}
\label{sec:results_xela}

Xe+La lies between Ar+Sc and Pb+Pb in size and is where the models separate. The
preliminary $4\pi$ data~\cite{Panova:2026} (Table~\ref{tab:xela_digitized}) are flat
between 30$A$ and 150$A$~GeV at $\avg{K^+}/\avg{\pi^+} \approx 0.16$. The core--corona
model ($\fcore \approx 0.72$) predicts a nearly flat excitation function at 0.19, the fitted
SMES a horn at 0.22 falling to 0.17, i.e., 15\% above the core--corona value at 30$A$ and
8\% below it at 75$A$--150$A$~GeV, and PHSD a horn with a 17\% maximum at 30$A$~GeV.
The final Xe+La spectra, including the lower energies, will therefore test directly
whether a horn exists in a system of this size, as SMES and PHSD predict, or not, as
the core--corona picture and the Ar+Sc data suggest.

\begin{table}
\caption{\label{tab:xela_digitized}
Digitized $\avg{K^+}/\avg{\pi^+}$ ($4\pi$)
for central Xe+La, from Fig.~5 of Ref.~\cite{Panova:2026}
(preliminary).}
\begin{ruledtabular}
\begin{tabular}{cccc}
$p_{\mathrm{lab}}$ ($A$ GeV/$c$) & $\sqrts$ (GeV) & $K^+/\pi^+$ & $\sigma_{\mathrm{tot}}$ \\
\hline
30  & 7.62  & 0.155 & 0.016 \\
40  & 8.76  & 0.161 & 0.016 \\
75  & 11.94 & 0.160 & 0.011 \\
150 & 16.84 & 0.160 & 0.014 \\
\end{tabular}
\end{ruledtabular}
\end{table}

\section{$K^-/\pi^-$ and the double ratio}
\label{sec:kminus}
\subsection{$K^-/\pi^-$}
\label{sec:results_kminus}

\begin{figure}
\includegraphics[width=\columnwidth]{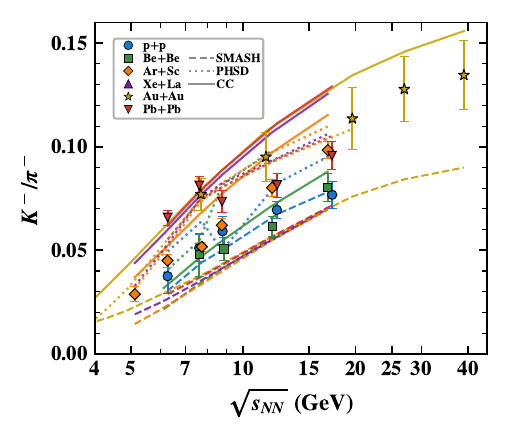}
\caption{\label{fig:kminus_piminus}
Mid-rapidity $K^-/\pi^-$ as a function of $\sqrts$ for p+p, Be+Be, Ar+Sc,
Pb+Pb and STAR Au+Au (model curves also for Xe+La). Dashed lines: SMASH; dotted lines:
PHSD; solid lines: core--corona.}
\end{figure}

The $K^-/\pi^-$ ratio (Fig.~\ref{fig:kminus_piminus}) rises with $\sqrts$ for all systems,
within the uncertainties, and shows no horn. The system-size ordering persists: at
$\sqrts \approx 7.6$~GeV, $K^-/\pi^-$ at mid-rapidity rises from 0.051 in p+p to 0.081 in
Pb+Pb. A larger baryon chemical potential suppresses $K^-$ ($\bar u s$) through the
strangeness chemical potential, so this rise is a conservative indication of enhanced
strangeness production in the larger systems. Among the models
(Table~\ref{tab:kminus_chi2}), PHSD comes closest for Pb+Pb ($\chindf = 3.9$ at
mid-rapidity, 1.1 in $4\pi$) and describes STAR Au+Au, SMASH is adequate for p+p and
Be+Be, and no model describes Ar+Sc at mid-rapidity ($\chindf \geq 19.8$).
The core--corona model improves Pb+Pb over SMASH much less for $K^-/\pi^-$
($53 \to 12$) than for $K^+/\pi^+$ ($38.6 \to 2.6$), because its core is an equilibrium gas that does not reproduce the measured
$K^+/K^-$ asymmetry (Secs.~\ref{sec:fits} and~\ref{sec:results_doubleratio}).

\begin{table}
\caption{\label{tab:kminus_chi2}
$\chindf$ (ndf $= n$) for $K^-/\pi^-$. $n$ is the number of data points (PHSD: 3 for
Be+Be at mid-rapidity, 4 in $4\pi$, and 3 for STAR Au+Au).}
\begin{ruledtabular}
\begin{tabular}{llcccc}
Acceptance & System & $n$ & SMASH & PHSD & CC \\
\hline
Mid-rapidity & p+p    & 5 & 0.78 & 2.57 & 0.78 \\
       & Be+Be  & 4 & 2.11 & 23.1 & 1.52 \\
       & Ar+Sc  & 6 & 46.5 & 22.1 & 19.8 \\
       & Pb+Pb  & 5 & 53.1 & 3.89 & 11.9 \\
       & Au+Au (STAR) & 5 & 11.8 & 0.09 & 1.23 \\
\hline
$4\pi$ & p+p    & 5 & 1.28 & 5.87 & 1.28 \\
       & Be+Be  & 5 & 0.67 & 4.74 & 2.65 \\
       & Ar+Sc  & 6 & 12.6 & 2.30 & 13.5 \\
       & Pb+Pb  & 5 & 46.4 & 1.12 & 20.1 \\
\end{tabular}
\end{ruledtabular}
\end{table}

\subsection{The double ratio}
\label{sec:results_doubleratio}

\begin{figure}
\includegraphics[width=\columnwidth]{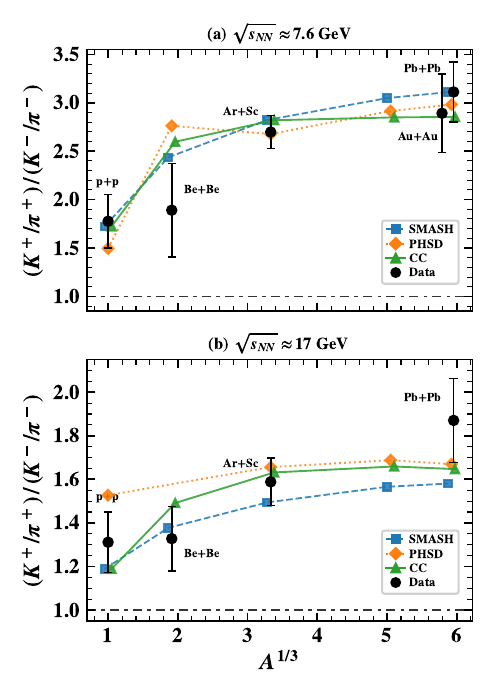}
\caption{\label{fig:double_ratio_syssize}
Double ratio $\mathcal{R}$ at mid-rapidity as a function of system size ($A^{1/3}$)
at (a)~$\sqrts \approx 7.6$~GeV and (b)~$\approx 17$~GeV.
Data (black circles), SMASH (dashed, squares), PHSD (dotted, diamonds),
and core--corona (solid, triangles).}
\end{figure}

The double ratio
\begin{equation}
  \mathcal{R} \;\equiv\; \frac{K^+/\pi^+}{K^-/\pi^-}
  \label{eq:double_ratio}
\end{equation}
measures the $K^+/K^-$ asymmetry. In a statistical source it is set by $\mu_S/T$ and
therefore by $\muB/T$ and $\gq$, but not by $\gs$ (Sec.~\ref{sec:shm_ratios}); in a
microscopic picture it reflects associated production ($NN \to N\Lambda K^+$), which
populates $K^+$ but not $K^-$. At $\sqrts \approx 7.6$~GeV the mid-rapidity data show
a step from $\mathcal{R} = 1.80 \pm 0.24$ (p+p and Be+Be, weighted mean) to
$2.79 \pm 0.15$ (Ar+Sc and Pb+Pb), a difference of $3.5\sigma$; at top SPS energy the
step is smaller (Fig.~\ref{fig:double_ratio_syssize}). At 7.6~GeV this step is
reproduced by all three models, including purely hadronic SMASH (1.72, 2.82 and 3.11 for
p+p, Ar+Sc and Pb+Pb against measured 1.78, 2.70 and 3.11); only Be+Be lies below all
models. The step in $\mathcal{R}$ therefore reflects the growth of baryon stopping and
associated production with system size and is not, by itself, evidence for
strangeness enhancement.

The $4\pi$ data add a quantitative constraint on hadronization.
Table~\ref{tab:double_ratio_shm} compares the measured $\mathcal{R}$ with the
statistical hadronization fits of Sec.~\ref{sec:fits}. The measured $\mathcal{R}$
exceeds the value of a single thermal source with $\gq = 1$ by 21--30\% in Pb+Pb and
12--32\% in Ar+Sc ($1.4$--$2.2\sigma$ and $1.5$--$3.1\sigma$ of the data uncertainty).
At the best $\gq$ the excess is reduced to 10--17\% ($0.8$--$1.4\sigma$) in Pb+Pb and
4--20\% ($0.6$--$2.1\sigma$) in Ar+Sc. In Pb+Pb the measured $\mathcal{R}$ is therefore
compatible with a single non-equilibrium thermal source and does not require an
additional $K^+$ source. In Ar+Sc the excess persists at 30$A$--75$A$~GeV
($1.5$--$2.1\sigma$), and no single source describes the yields
(Sec.~\ref{sec:fits_small}); a $K^+/K^-$ asymmetry beyond the thermal value at the
baryon content fixed by the other hadrons is expected if a part of the $K^+$ yield comes
from associated production in nucleon--nucleon-like collisions, i.e., from a corona.
In Be+Be the fits reproduce $\mathcal{R}$.

\begin{table}
\caption{\label{tab:double_ratio_shm}
$4\pi$ double ratio $\mathcal{R}$: data and statistical hadronization fits at
$\gq = 1$ and at the best $\gq$ (Table~\ref{tab:shm_fits}). $E_{\mathrm{beam}}$ in
$A$~GeV.}
\begin{ruledtabular}
\begin{tabular}{lcccc}
System & $E_{\mathrm{beam}}$ & Data & $\gq = 1$ & best $\gq$ \\
\hline
Pb+Pb & 30  & $3.76 \pm 0.38$ & 3.07 & 3.42 \\
      & 40  & $3.42 \pm 0.38$ & 2.63 & 2.98 \\
      & 80  & $2.61 \pm 0.28$ & 2.01 & 2.23 \\
      & 158 & $2.13 \pm 0.27$ & 1.76 & 1.91 \\
Ar+Sc & 30  & $3.13 \pm 0.24$ & 2.38 & 2.62 \\
      & 40  & $2.76 \pm 0.24$ & 2.16 & 2.38 \\
      & 75  & $2.27 \pm 0.15$ & 1.89 & 2.05 \\
      & 150 & $1.85 \pm 0.13$ & 1.65 & 1.78 \\
Be+Be & 40  & $2.47 \pm 0.60$ & 2.15 & -- \\
      & 75  & $1.89 \pm 0.37$ & 1.92 & -- \\
      & 150 & $1.63 \pm 0.30$ & 1.70 & -- \\
\end{tabular}
\end{ruledtabular}
\end{table}

\section{Discussion: the horn and the dynamics of hadronization}
\label{sec:discussion}

\paragraph*{Where the horn is and where it is not.}
The per-participant decomposition separates the two ingredients of the horn along the
system-size axis. The ``onset of fireball'' between Be+Be and Ar+Sc is an onset in
strangeness production per participant, consistent with strangeness production that is
amplified by multiple collisions per participant~\cite{Rafelski:1982pu,Koch:1986ud}, and
with the geometric core fraction of the core--corona model. The horn, in contrast, is a
feature of the largest system only, and it originates on the entropy side: a pion yield
per participant that is suppressed at 20$A$--40$A$~GeV and then rises faster than in
Ar+Sc, while strangeness per participant grows at a similar rate in both. A rise of
strangeness production that is overtaken by a faster rise of entropy production is
common to the two pictures of Sec.~\ref{sec:intro}; they differ in how the entropy
excess reaches the hadrons. In SMES~\cite{Gazdzicki:1998vd} it is an equilibrium
property of a two-phase equation of state with a mixed phase. In the kinetic
picture~\cite{Rafelski:1982pu,Koch:1986ud,Letessier:2002book}, strangeness produced by
gluon fusion and amplified by multiple collisions is overtaken by the entropy excess of
the deconfined phase, which sudden hadronization carries into the hadrons with
$\gq > 1$~\cite{Rafelski:2000by,Letessier:2008}.

\paragraph*{Chemical non-equilibrium in Pb+Pb.}
The statistical hadronization fits express the same observation in terms of
hadronization. From 30$A$~GeV upward, the Pb+Pb yields prefer a light-quark phase space
at the pion condensation limit, $\gq \simeq e^{m_\pi/2T}$, at $T \approx 129$--136~MeV,
over chemical equilibrium at $T \approx 144$--151~MeV; at $\gq = 1$ the pion and hyperon
yields are underestimated together. At 20$A$~GeV the data do not constrain $\gq$, so
the present data and analysis do not inform on the hadronization below 30$A$~GeV: they
establish a preference for $\gq > 1$ from 30$A$~GeV, the maximum of
$\avg{K^+}/\avg{\pi^+}$, upward, but not a change between 20$A$ and 30$A$~GeV. Such a
change of hadronization dynamics between 20$A$ and 30$A$~GeV was reported in
Ref.~\cite{Rafelski:2008av} (see also Ref.~\cite{Letessier:2008}). The preference from
30$A$~GeV upward is what is expected for sudden hadronization of an entropy-rich source,
in which the excess entropy of the deconfined phase is carried into the hadron yields by
an over-saturated light-quark phase space instead of by a long-lived mixed phase or by
reheating~\cite{Letessier:1999rm,Rafelski:2000by,Letessier:2008}. The
robustness tests of Sec.~\ref{sec:fits_robust} show this excess directly: at
30$A$--80$A$~GeV the produced hadrons are compatible with equilibrium among themselves,
but not with the net-baryon number, and the pion part of the preference exists only
through Bose--Einstein enhancement near the condensation limit. At 158$A$~GeV the
preference is fragile: it disappears without the pion yields or without $K^-$. A
long-lived mixed phase would imply a long emission duration, which pion interferometry
does not show across the same energy
range~\cite{Rischke:1996em,Lisa:2005dd,NA49:2007fqa}.
The SMES variants reproduce the heavy-system excitation function through a mixed phase and
are subject to this constraint; we regard their description of Pb+Pb and Au+Au as
phenomenological.

\paragraph*{Intermediate and small systems.}
Ar+Sc has strangeness per participant close to Pb+Pb, yet its
$\avg{K^+}/\avg{\pi^+}$ rises monotonically with energy. The statistical hadronization
fits locate the difference: in Ar+Sc the pion yields are described at $\gq = 1$, the
preference for $\gq > 1$ comes entirely from the $K^+/K^-$ asymmetry at the baryon
chemical potential fixed by $\avg{W}$ and $\bar p$ (Sec.~\ref{sec:fits_robust}), and no
single source describes the data. An excess of $K^+$ over $K^-$ beyond thermal expectations is
the signature of associated production in nucleon--nucleon-like collisions, i.e., of a
corona; the core--corona model assigns about 40\% of the Ar+Sc participants to the corona.
The per-participant data point the same way: Ar+Sc has the largest pion yield per
participant of all systems at 19$A$--40$A$~GeV, with no sign of the low-energy
suppression seen in Pb+Pb. Whether this suppression in Pb+Pb reflects the entropy
balance of a deconfined source, or hadronic effects of the larger baryon density such as
pion absorption in baryon resonances, cannot be decided from yields alone. In either
case the horn appears only where a single hadronizing source dominates and its entropy
production changes rapidly with energy. For Be+Be the available multiplicities do not
discriminate between $\gq = 1$ and $\gq > 1$; the robust quantity is
$\gs/\gq \approx 0.4$, well below the Ar+Sc values.

\paragraph*{Transport.}
PHSD with chiral symmetry restoration reproduces the horn in central Pb+Pb and Au+Au,
which shows that a non-equilibrium microscopic mechanism can generate it without a
first-order transition~\cite{Cassing:2016,Palmese:2016rtq}. In PHSD, however, the horn
appears also in Ar+Sc and Xe+La, where the data show none, and the
mid-rapidity kaon-to-pion ratios of Ar+Sc and Be+Be are overpredicted. The system-size
axis thus constrains the mechanism that produces the horn in this approach. Purely
hadronic transport (SMASH) produces neither the horn nor the strangeness step, but at
the horn energy it reproduces the system-size dependence of the $K^+/K^-$ asymmetry.

\paragraph*{Relation to the LHC.}
At the LHC the system-size question takes a different form. ALICE finds that the
yields of strange and multi-strange hadrons relative to pions rise smoothly with the
charged-particle multiplicity from p+p through p+Pb to Pb+Pb, independent of the
collision system at the same multiplicity~\cite{ALICE:2017}; no threshold in system size
analogous to the SPS step between Be+Be and Ar+Sc is seen. For the 20\% most central
Pb+Pb collisions at $\sqrts = 2.76$~TeV, Petr\'an \textit{et al.}\ obtain $\chindf = 9.5/9$ with
$\gq = 1.63$, $\gs = 2.08$ and $T = 138$~MeV, and 64/11 in full
equilibrium~\cite{Petran:2013lja}; for $\gq = 1$ with free $\gs$, the parameter table in
Fig.~1(a) of Ref.~\cite{Petran:2013lja} gives 35/10. At all centralities except the most
peripheral one, $\chi^2$ decreases toward $\gq$ at the condensation limit, as we find for
Pb+Pb from 30$A$~GeV upward. If this is the signature of sudden hadronization of an
entropy-rich source, it is present from the horn energy upward and persists to the LHC, while
the small SPS systems, in which a corona contributes substantially, do not show it.

\paragraph*{What kaon-to-pion ratios can and cannot show.}
Because $\avg{K^+}/\avg{\pi^+}$ fixes only $\gs/\gq$, a strangeness saturation factor
extracted from it with $\gq = 1$ measures $\gs/\gq$: about 0.4 in Be+Be, 0.5--0.8 in
Ar+Sc and 0.9--1.0 in Pb+Pb at the best-fit $\gq$. The value near unity in Pb+Pb is
reached with $\gs \approx \gq \approx 1.6$ and does not indicate strangeness chemical
equilibrium. The double ratio probes $\muB/T$ and $\gq$, and its step with system size
is reproduced by hadronic transport. Chemical equilibrium, where it applies, has to emerge
from fits in which $\gq$ and $\gs$ are free, rather than be assumed.

\begin{table*}
\caption{\label{tab:fit_tables}
Measured $4\pi$ multiplicities (Table~\ref{tab:multiplicities}) with total
uncertainty $\sigma$, and statistical hadronization values with pulls
$(N^{\mathrm{fit}} - N^{\mathrm{exp}})/\sigma$ at $\gq = 1$ and at the minimum of the
$\chi^2$ profile in $\gq$; all other parameters as in Table~\ref{tab:shm_fits}.
$\avg{W}$ enters as the net-baryon number.}
\begin{ruledtabular}
\begin{tabular}{lcc|cc|cc}
 & & & \multicolumn{2}{c|}{$\gq = 1$} & \multicolumn{2}{c}{$\gq$ at $\chi^2$ minimum} \\
Hadron & Data & $\sigma$ & Fit & Pull & Fit & Pull \\
\hline
\multicolumn{7}{l}{\textit{Pb+Pb, 30$A$~GeV}: $\gq = 1$ ($\chi^2 = 14.2$) and $\gq = 1.67$ ($\chi^2 = 3.6$)} \\
$\avg{W}$ & 349 & 5.10 & 352 & $+0.6$ & 350 & $+0.1$ \\
$\pi^-$ & 274 & 14.0 & 259 & $-1.1$ & 274 & $0.0$ \\
$\pi^+$ & 241 & 12.0 & 230 & $-0.9$ & 241 & $0.0$ \\
$K^+$ & 52.9 & 3.37 & 44.9 & $-2.4$ & 48.4 & $-1.3$ \\
$K^-$ & 16.0 & 0.447 & 16.5 & $+1.0$ & 16.1 & $+0.1$ \\
$\phi$ & 1.84 & 0.364 & 2.35 & $+1.4$ & 2.25 & $+1.1$ \\
$\Lambda$ & 36.9 & 3.31 & 33.3 & $-1.1$ & 36.3 & $-0.2$ \\
$\bar\Lambda$ & 0.390 & 0.045 & 0.393 & $+0.1$ & 0.392 & $0.0$ \\
$\Xi^-$ & 2.42 & 0.347 & 2.08 & $-1.0$ & 2.56 & $+0.4$ \\
$\bar\Xi^+$ & 0.120 & 0.036 & 0.084 & $-1.0$ & 0.099 & $-0.6$ \\
\hline
\multicolumn{7}{l}{\textit{Pb+Pb, 158$A$~GeV}: $\gq = 1$ ($\chi^2 = 21.3$) and $\gq = 1.64$ ($\chi^2 = 14.5$)} \\
$\avg{W}$ & 362 & 5.10 & 364 & $+0.3$ & 363 & $+0.1$ \\
$\pi^-$ & 639 & 35.4 & 577 & $-1.8$ & 600 & $-1.1$ \\
$\pi^+$ & 594 & 33.1 & 544 & $-1.5$ & 564 & $-0.9$ \\
$K^+$ & 103 & 7.07 & 102 & $-0.2$ & 105 & $+0.3$ \\
$K^-$ & 51.9 & 3.55 & 61.3 & $+2.7$ & 58.4 & $+1.8$ \\
$\phi$ & 8.46 & 0.503 & 8.07 & $-0.8$ & 7.63 & $-1.7$ \\
$\Lambda$ & 48.5 & 9.00 & 48.1 & $0.0$ & 51.9 & $+0.4$ \\
$\bar\Lambda$ & 3.32 & 0.380 & 3.77 & $+1.2$ & 3.66 & $+0.9$ \\
$\Xi^-$ & 4.37 & 0.676 & 4.30 & $-0.1$ & 5.24 & $+1.3$ \\
$\bar\Xi^+$ & 0.713 & 0.103 & 0.627 & $-0.8$ & 0.723 & $+0.1$ \\
$\Omega^-$ & 0.594 & 0.158 & 0.389 & $-1.3$ & 0.481 & $-0.7$ \\
$\bar\Omega^+$ & 0.263 & 0.073 & 0.109 & $-2.1$ & 0.132 & $-1.8$ \\
\hline
\multicolumn{7}{l}{\textit{Ar+Sc, 30$A$~GeV}: $\gq = 1$ ($\chi^2 = 28.7$) and $\gq = 1.62$ ($\chi^2 = 13.4$)} \\
$\avg{W}$ & 60.6 & 1.20 & 61.8 & $+1.0$ & 61.5 & $+0.8$ \\
$\pi^+$ & 58.2 & 2.34 & 55.5 & $-1.1$ & 56.1 & $-0.9$ \\
$\pi^-$ & 60.0 & 2.34 & 57.7 & $-1.0$ & 58.5 & $-0.7$ \\
$K^+$ & 7.44 & 0.313 & 6.10 & $-4.3$ & 6.55 & $-2.8$ \\
$K^-$ & 2.45 & 0.082 & 2.66 & $+2.6$ & 2.60 & $+1.9$ \\
$\bar p$ & 0.116 & 0.014 & 0.113 & $-0.2$ & 0.114 & $-0.1$ \\
\end{tabular}
\end{ruledtabular}
\end{table*}

\section{Conclusions}
\label{sec:conclusions}

We have tested the standard interpretation of the $K^+/\pi^+$ horn, and the
chemical-equilibrium assumption often used to analyze it, on the system-size axis
provided by NA61/SHINE and NA49. The Pb+Pb fits confirm earlier work with the final data;
the Be+Be and Ar+Sc fits (item~3) are new. Our findings are:

\begin{enumerate}

\item \textbf{Strangeness step and entropy horn.}
Between Be+Be and Ar+Sc, strangeness per wounded nucleon increases by a factor
1.8--2.3 at every SPS energy, at nearly unchanged pion yield per participant: the
system-size step is a step in strangeness production. The horn is present only in
Pb+Pb, and it originates in the pion yield per participant, which is suppressed at
20$A$--40$A$~GeV and then rises faster than in Ar+Sc.

\item \textbf{Chemical non-equilibrium in central Pb+Pb: confirmation with final data.}
With the final NA49 data and the full hadron spectrum, we confirm the result of
Letessier, Rafelski, and collaborators: chemical equilibrium of light quarks is
disfavored from 30$A$~GeV, the maximum of the horn, upward ($\Delta\chi^2 = 7$--15),
with $\gq$ at the pion condensation limit, $T \approx 129$--136~MeV and
$\gs \approx 1.4$--1.7; at 20$A$~GeV $\gq$ is not constrained. The preference is
independent of the hadron list and the data compilation and, at 30$A$--80$A$~GeV, of any
single hadron yield. It rests on the two ingredients of the entropy-excess picture: an
excess of produced hadrons per net baryon~\cite{Letessier:1992xd} and, for the pions,
Bose--Einstein enhancement near the condensation limit~\cite{Letessier:1999rm}. The change of hadronization dynamics between 20$A$ and 30$A$~GeV
reported in Ref.~\cite{Rafelski:2008av} is not tested by the present data, which do not
constrain $\gq$ below 30$A$~GeV.

\item \textbf{Ar+Sc and Be+Be: new non-equilibrium fits.}
The NA61/SHINE multiplicities of Be+Be and Ar+Sc are fitted here with $\gq$ free. In Ar+Sc, $\gq = 1$ is disfavored as well, but through the $K^+/K^-$ asymmetry rather
than the pion yields, and no single thermal source describes the data, which points to a
substantial corona. The Be+Be data are compatible with equilibrium but do not
discriminate; they fix $\gs/\gq \approx 0.4$, against 0.5--0.8 in Ar+Sc with exact
strangeness conservation included.

\item \textbf{Models.}
Hadronic transport produces neither the horn nor the strangeness step. PHSD with chiral
symmetry restoration reproduces the Pb+Pb and Au+Au horn but predicts one in Ar+Sc and
Xe+La, and overpredicts the mid-rapidity Ar+Sc and Be+Be ratios. The core--corona model,
with one parameter calibrated on Pb+Pb, improves on SMASH for Ar+Sc, STAR Au+Au and the
preliminary Xe+La data, but overshoots Be+Be and fails at AGS energies.
SMES describes the heavy systems through a mixed phase and overpredicts Ar+Sc.

\item \textbf{Observables.}
$\avg{K^+}/\avg{\pi^+}$ constrains only $\gs/\gq$, and the double ratio
$(K^+/\pi^+)/(K^-/\pi^-)$ probes $\muB/T$ and $\gq$, not strangeness saturation. The
step of the double ratio with system size is reproduced by hadronic transport. Its
$4\pi$ value exceeds single-source statistical hadronization at $\gq = 1$ in Pb+Pb and
Ar+Sc (by 1.4--3.1$\sigma$). At the best $\gq$ the Pb+Pb excess (0.8--1.4$\sigma$) is
not significant, whereas in Ar+Sc an excess of up to 2.1$\sigma$ remains, consistent
with a corona contribution to $K^+$.

\end{enumerate}

\paragraph*{Outlook.}
The final NA61/SHINE Xe+La spectra, including the energies below 30$A$~GeV, will test
whether a horn exists in a system between Ar+Sc and Pb+Pb, as PHSD and SMES predict.
Multiplicities of $\Lambda$, $\Xi$ and $\phi$ in Be+Be and Ar+Sc would break the
degeneracy between $T$ and $\gq$ in the small systems and decide whether chemical
non-equilibrium of light quarks sets in between Be+Be and Pb+Pb; a light-ion run at
the SPS (e.g., O+O) would place a system in the transition region. The STAR beam energy
scan~II and its fixed-target program at $\sqrts = 3$--7.7~GeV, CBM at FAIR~\cite{Ablyazimov:2017guv},
and MPD at NICA will extend the Au+Au excitation function and the system-size dependence
at high baryon density.

\begin{acknowledgments}
We thank the anonymous referee, whose outline of the analysis this revised work follows,
for detailed and constructive comments. We thank the PHSD team for valuable advice and
guidance on the PHSD calculations. We acknowledge
computational resources provided by the Indian Institute of Technology Mandi. We thank the
NA49, NA61/SHINE, and STAR Collaborations for making their data publicly available through
HEPData.
Claude Opus~5.5 (Anthropic) assisted with the literature review, parts of the analysis
code, and the bibliography.
\end{acknowledgments}

\section*{Data Availability}
All experimental data used in this analysis are compiled from published sources cited in
Sec.~\ref{sec:data}. The analysis code is available from the corresponding author upon
reasonable request.

\appendix

\section{Measured and fitted multiplicities}
\label{app:fit_tables}

Table~\ref{tab:fit_tables} lists, for three representative cases, the measured $4\pi$
multiplicities and the statistical hadronization values at $\gq = 1$ and at the minimum
of the $\chi^2$ profile in $\gq$ (Table~\ref{tab:shm_fits}), with the pulls
$(N_i^{\mathrm{fit}} - N_i^{\mathrm{exp}})/\sigma_i$. In Pb+Pb at 30$A$~GeV the
non-equilibrium fit restores the pion, $K^+$, $\Lambda$ and $\Xi^-$ yields together; at
158$A$~GeV it improves the pions and $K^-$ at the expense of $\phi$. In Ar+Sc the pions
are described at $\gq = 1$ and the residual tension is carried by $K^+$ and $K^-$
(Sec.~\ref{sec:fits_small}).

\bibliographystyle{apsrev4-2}
\bibliography{strangeness_horn}

\end{document}